\documentclass{article}

\usepackage{parskip}
\usepackage[a4paper, total={6in, 8in}]{geometry}
\usepackage{listings}
\usepackage{amssymb}
\usepackage{mathtools}

\newcommand{\bang}[1]{\operatorname{! #1}}
\newcommand{\one}{\textbf{1}}
\newcommand{\zero}{\textbf{0}}
\newcommand{\mgu}[2]{\mathrm{mgu}(#1, #2)}
\newcommand{\mguto}[2]{\mathrm{mguto}(#1, #2)}
\newcommand{\newvar}{\mathit{newvar}}
\newcommand{\inst}[1]{\mathit{inst}(#1)}
\newcommand{\letin}[2]{\textsf{let } #1 \textsf{ in } #2}
\newcommand{\unit}{\textsf{tt}}
\newcommand{\fst}{\textsf{fst} ~}
\newcommand{\snd}{\textsf{snd} ~}
\newcommand{\inl}{\textsf{inl} ~}
\newcommand{\inr}{\textsf{inr} ~}
\newcommand{\caseof}[5]{\textsf{case} ~ #1 ~ \textsf{of inl} ~ #2 \Rightarrow #3 ~ | ~ \textsf{inr} ~ #4 \Rightarrow #5}
\newcommand{\absurd}{\textsf{absurd} ~ }
\newcommand{\fold}[2]{\textsf{fold}^{#1} #2}
\newcommand{\unfoldi}{\textsf{unfold} ~}
\newcommand{\unfold}[2]{\textsf{unfold}^{#1} #2}
\newcommand{\fix}[2]{\textsf{fix} ~ #1. #2}

\title{An Affine Type System with Hindley-Milner Style Type Inference}
\author{Gonglin Li}
\date{March 31, 2022}

\begin{document}

\maketitle

\begin{abstract}
This article first provides an algorithm W based type inference algorithm for an affine type system. Then the article further assumes the language equipped with the above type system uses lazy evaluation, and explores the possibility of representing the !-modality as a user-defined type synonym with the power of the newly gained polymorphism.
\end{abstract}

\section{Affine Type System}

Affine type system is the type theory counterpart of affine logic, a variant of linear logic, where every proposition may be ``used'' at most once \cite{hoffmann19}. Affine logic doesn't accept the contraction rule:
\begin{displaymath}
  \frac{\Gamma, A, A\vdash B}{\Gamma, A \vdash B}
\end{displaymath}
In an affine type system, locally bound variables may be viewed as resources, and will be consumed after use. This lets the compiler determine when to deallocate their memory, as is done in the programming language Rust.

Apart from the commonly known product and sum type, which is referred to as \textit{tensor} and \textit{plus} respectively, affine type system has a third binary connective, namely ``with''. Speaking roughly, \textit{a with b} means you can choose from \textit{a} and \textit{b}, but you can't have both. \textbf{1} is the unit for \textit{with}.

There is also a type representing constant truth in affine logic, which allows the contraction rule, namely the !-modality. It corresponds to globally defined terms. These terms are usually compiled to static code and have nothing to do with memory (de)allocation.

Unlike in a linear type system, where each variable must be used exactly once, in affine type systems a variable is not guaranteed to be used. Thus we intend this type system be equipped in a lazy language, to avoid unnecessary evaluation, though nothing has to do with this assumption until section 4.

\bigskip

In this article we will stick to the following notions to avoid confusion:
\begin{itemize}
  \item $\otimes$, denoting tensor
  \item $\oplus$, denoting plus
  \item $\&$, denoting with
  \item $!$, written as a prefix, denoting the !-modality
  \item $\multimap$, denoting linear implication
\end{itemize}

\section{Introduction}

In Hindley-Milner type system, there is no need for predefined product and sum type. We may encode them using polymorphic functions, as shown in Haskell:
\begin{lstlisting}
pair :: a -> b -> (a -> b -> c) -> c
pair x y f = f x y

fst :: ((a -> b -> a) -> c) -> c
fst p = p (\x y -> x)
\end{lstlisting}
However, for \lstinline{fst} to have the ``correct'' type (i.e. to match with \lstinline{pair}), we would have to introduce rank-2 type:
\begin{lstlisting}
fst :: (forall c. (a -> b -> c) -> c) -> a
fst p = p (\x y -> x)
\end{lstlisting}
which already exceeds the expressive power of HM. Instead, we could introduce them as type constants, and define their constructors and elemintors manually.
\begin{lstlisting}
-- implementation hidden
data Pair :: * -> * -> *
constructPair :: a -> b -> Pair a b
elimPair :: (a -> b -> c) -> Pair a b -> c

fst :: Pair a b -> a
fst = elimPair (\x y -> x)
\end{lstlisting}

But this trick doesn't apply to affine type system, because the constructor of \textit{with} cannot even be expressed! Therefore, we have to introduce new typing rules for these types, and generalize algorithm W for a type inference algorithm.

\section{Typing Rules}

We'll use judgements of the following form as introduced in \cite[section~6.2]{pfenning02}:
\begin{displaymath}
  \Gamma \backslash \Delta \vdash e : \tau, S
\end{displaymath}
where $\Gamma$ is the linear context available for typechecking $e$ and $\Delta \subseteq \Gamma$ contains the (maybe refined) linear hypotheses not used in $e$. Thus when dealing with a rule with two or more premises, instead of guessing how to split the context, we may give the whole context to the first one, typecheck the first subterm, and pass the remaining context to the second premise. We don't bother to introduce the non-linear context; they could be represented by $!-$ types.

$S$ is the substitution yield by unifications in the process of type inference. Substitutions are defined as maps from type variables to types, and are sometimes considered as functions, thus we may use the same notion as functions by the abuse of notion. Their composition is defined in the following means: applying $S_1 S_2$ to $\tau$ is the same as applying $S_1$ to $S_2 \tau$, i.e. in the normal composition order.

\subsection{Linear Implication}

For linear implication introduction (i.e. lambda abstraction), we must make sure the introduced variable doesn't go out of scope:

\begin{displaymath}
  \frac
    {\sigma \leftarrow \newvar \qquad \Gamma, w : \sigma \backslash \Delta \vdash e : \tau, S}
    {\Gamma \backslash \Delta - w \vdash \lambda w. e : S \sigma \multimap \tau, S} 
  \multimap \mathrm{I}
\end{displaymath}

$\newvar$ is rendered in italic, indicating it has side effect, but it's only limited in creating fresh type variables.

For linear implication elimination (i.e. application), we would need to unify two types:

\begin{displaymath}
  \frac
    {\begin{array}{ll}
      \Gamma \backslash \Delta \vdash e_0 : \tau_0, S_0 \qquad &
      S_0 \Delta \backslash \Theta \vdash e_1 : \tau_1, S_1 \\
      \tau' \leftarrow \newvar \qquad &
      S_2 = \mgu{S_1 \tau_0}{\tau_1 \multimap \tau'}
     \end{array}}
    {\Gamma \backslash \Theta \vdash e_0 e_1 : S_2 \tau', S_2 S_1 S_0} 
  \multimap \mathrm{E}
\end{displaymath}

where $\mathrm{mgu}$ is a side-effect free algorithm that determines the most general unifier of two types. This algorithm may fail, in which case type inference also fails. An implementation in Haskell is included in the Appendix.

\subsection{Let-generalization and Instantiation}

The most important observation when combining HM with affine type system is that explicit \textsf{let} construct is no longer necessary. In HM, let-generalization is only needed when a let-bound variable is used more than once, and it has been instantialized to have different types; however in an affine type system this becomes impossible.

\begin{displaymath}
\displaystyle \frac
  { \displaystyle \frac 
      { \displaystyle \frac 
          {} 
          {\Gamma \vdash id : \mathrm{Int} \to \mathrm{Int}} \qquad
        \displaystyle \frac 
          {}
          {\Gamma \vdash \textrm{3} : \mathrm{Int}}
      } 
      {\Gamma \vdash id \textrm{ 3} : \mathrm{Int}} \qquad
    \displaystyle \frac 
      { \displaystyle \frac 
          {} 
          {\Gamma \vdash id : \mathrm{Bool} \to \mathrm{Bool}} \qquad
        \displaystyle \frac 
          {}
          {\Gamma \vdash \textrm{True} : \mathrm{Bool}}
      } 
      {\Gamma \vdash id \textrm{ True} : \mathrm{Bool}}}
  {\Gamma \vdash (id \textrm{ 3}, id \textrm{ True}) : \mathrm{Int} \times \mathrm{Bool}}
\end{displaymath}
\centerline{(An example of using generalized variable twice, where $\Gamma = \cdot, id : \forall \tau. \tau \to \tau$.)}

\bigskip

Thus we would have a plain old stlc-style variable introduction rule:
\begin{displaymath}
  \frac
    {}
    {\Gamma, v : \tau \backslash \Gamma \vdash v : \tau, \emptyset}
  \mathrm{Var}
\end{displaymath}

Instantiation only occurs when we're trying to bring a globally defined term into scope, since they're implicitly bound by let expressions:
\begin{displaymath}
  \frac
    {E : \tau \qquad \tau' \leftarrow \inst{\tau}}
    {\Gamma \backslash \Gamma \vdash E : \tau', \emptyset}
  \mathrm{Intro}
\end{displaymath}
where $E : \tau$ is short for $\cdot \backslash \cdot \vdash E : \tau$. $\inst{\tau}$ substitutes all free variables in $\tau$ with fresh variables. We'll revisit this rule when we come across the !-modality.

\subsection{Tensor}

Because of the existence of currying, the introduction rule for tensor looks much like that of linear implication: the given context is passed to the first premise, and the remaining part is fed to the second one.

\begin{displaymath}
  \frac
    {\Gamma \backslash \Delta \vdash e_0 : \tau_0, S_0 \qquad
     S_0 \Delta \backslash \Theta \vdash e_1 : \tau_1, S_1}
    {\Gamma \backslash \Theta \vdash e_0 \otimes e_1 : S_1 \tau_0 \otimes \tau_1, S_1 S_0}
  \otimes \mathrm{I}
\end{displaymath}

\begin{displaymath}
  \frac
    {\begin{array}{ll}
      \Gamma \backslash \Delta \vdash e_0 : \tau_0, S_0 \qquad &
      \alpha, \beta \leftarrow \newvar \\
      S_1 = \mgu{\tau_0}{\alpha \otimes \beta} \qquad &
      S_0 \Delta, w_0 : S_1 \alpha, w_1 : S_1 \beta \backslash \Theta \vdash e_1 : \tau_1, S_2
    \end{array}}
    {\Gamma \backslash \Theta - w_0 - w_1 \vdash \letin{w_0 \otimes w_1 = e_0}{e_1} : \tau_1, S_2 S_1 S_0}
  \otimes \mathrm{E}
\end{displaymath}

\subsection{With}

The idea of \textit{with} is that if $\Gamma \vdash A$ and $\Gamma \vdash B$, then $\Gamma \vdash A \& B$; conversely, if $\Gamma, A \vdash C$ or $\Gamma, B \vdash C$ then $\Gamma, A \& B \vdash C$. \lstinline{a & b} may be viewed as \lstinline{forall c. Either (a 

\begin{displaymath}
  \frac
    {\Gamma \backslash \Delta \vdash e_0 : \tau_0, S_0 \qquad
     S_0 \Gamma \backslash \Theta \vdash e_1 : \tau_1, S_1}
    {\Gamma \backslash \Delta \cap \Theta \vdash e_0 \& e_1 : S_1 \tau_0 \& \tau_1, S_1 S_0}
  \&\mathrm{I}
\end{displaymath}

\begin{displaymath}
  \frac
    {\Gamma \backslash \Delta \vdash e : \tau, S_0 \qquad
    \alpha, \beta \leftarrow \newvar \qquad
    S_1 = \mgu{\tau}{\alpha \& \beta}}
    {\Gamma \backslash \Delta \vdash \fst e : S_1 \alpha, S_1 S_0}
  \&\mathrm{E_L}
\end{displaymath}

\begin{displaymath}
  \frac
    {\Gamma \backslash \Delta \vdash e : \tau, S_0 \qquad
    \alpha, \beta \leftarrow \newvar \qquad
    S_1 = \mgu{\tau}{\alpha \& \beta}}
    {\Gamma \backslash \Delta \vdash \snd e : S_1 \beta, S_1 S_0}
  \&\mathrm{E_R}
\end{displaymath}

\subsection{The \textbf{1} Type}

\textbf{1} is the type with only one value. It's the unit of tensor and with.

\begin{displaymath}
  \frac
    {}
    {\Gamma \backslash \Gamma \vdash \unit : \one, \emptyset}
  \one\mathrm{I}
\end{displaymath}

Though \textbf{1} has no elimination rule, it may be discarded anytime in an affine type system.

\subsection{Plus}

Plus is the dual of with. However, though its rules are essentially in correspondence with those of with, it has the most complex elimination rule.

\begin{displaymath}
  \frac
    {\Gamma \backslash \Delta \vdash e : \alpha, S \qquad 
    \beta \leftarrow \newvar}
    {\Gamma \backslash \Delta \vdash \inl e : \alpha \oplus \beta, S}
  \oplus \mathrm{I_L}
\end{displaymath}

\begin{displaymath}
  \frac
    {\Gamma \backslash \Delta \vdash e : \beta, S \qquad 
    \alpha \leftarrow \newvar}
    {\Gamma \backslash \Delta \vdash \inr e : \alpha \oplus \beta, S}
  \oplus \mathrm{I_R}
\end{displaymath}

\begin{displaymath}
  \frac
    {\begin{array}{ll}
      \Gamma \backslash \Delta \vdash e : \sigma, S_0 \qquad &
      \alpha, \beta \leftarrow \newvar \\
      S_1 = \mgu{\sigma}{\alpha \oplus \beta} \qquad &
      S_0 \Delta, w_0 : S_1 \alpha \backslash \Theta \vdash e_0 : \tau', S_2 \\
      S_2 (S_0 \Delta, w_1 : S_1 \beta) \backslash \Xi \vdash e_1 : \tau, S_3 \qquad &
      S_4 = \mgu{S_3 \tau'}{\tau}
    \end{array}}
    {\Gamma \backslash (\Theta - w_0) \cap (\Xi - w_1) \vdash \caseof{e}{w_0}{e_0}{w_1}{e_1} : S_4 \tau, S_4 S_3 S_2 S_1 S_0}
  \oplus \mathrm{E}
\end{displaymath}

\subsection{The \textbf{0} Type}

\textbf{0} is the type with no value. It's the unit of plus. It corresponds to falsehood, hence has no introduction rule.

\begin{displaymath}
  \frac
    {\Gamma \backslash \Delta \vdash e : \sigma, S_0 \qquad
    S_1 = \mgu{\sigma}{\zero} \qquad
    \tau \leftarrow \newvar}
    {\Gamma \backslash \Delta \vdash \absurd e : \tau, S_1 S_0}
  \zero \mathrm{E}
\end{displaymath}

\section{Recursive Types}

With our current type system, we're able to define finite types, e.g. $\mathrm{Bool} = \one \oplus \one$. However, to express infinite types such as $\mathrm{Nat}$, we'll have to enrich our type system. Thus we introduce \textbf{recursive types} \cite[section~6.4]{pfenning02}, written as $\mu \alpha. M$. Simply put, $\mu \alpha. M$ is the greatest fix point (since we're working with a lazy language) of the type-level lambda $\lambda (\alpha : *). M : * \to *$, where $*$ stands for the universe $\mathcal{U}_0$. By this definition, we may have $\mathrm{Nat} = \mu \alpha. \one \oplus \alpha$. However, instead of considering $\mu \alpha. M$ and $M[\alpha := \mu \alpha. M]$ equal, we'd rather make them isomorphic, and require explicit constructs to roll and unroll recursive types, that is, their introduction and elimination rule.

\begin{displaymath}
  \frac
    {\Gamma \backslash \Delta \vdash e : M[\alpha := \mu \alpha. M]}
    {\Gamma \backslash \Delta \vdash \fold{\mu \alpha. M}{e} : \mu \alpha. M}
  \mu \mathrm{I}
  \qquad
  \frac
    {\Gamma \backslash \Delta \vdash e : \mu \alpha. M}
    {\Gamma \backslash \Delta \vdash \unfoldi e : M[\alpha := \mu \alpha. M]}
  \mu \mathrm{E}
\end{displaymath}

Above are the stlc style typing rules for recursive types, where \textsf{fold} needs a type annotation while \textsf{unfold} not. Now we generalize these rules for a version with type inference.

We won't try to infer the type argument of \textsf{fold}: this would require a unification involving $\forall (f : * \to *). f (\mu \alpha. f \alpha)$, which is second order and is generally undecidable. Thus we have the HM version of $\mu \mathrm{I}$:
\begin{displaymath}
  \frac
    {\Gamma \backslash \Delta \vdash e : \tau, S_0 \qquad S_1 = \mguto{\tau}{M[\alpha := \mu \alpha. M]}}
    {\Gamma \backslash \Delta \vdash \fold{\mu \alpha. M}{e} : S_1(\mu \alpha. M), S_1 S_0}
  \mu \mathrm{I}
\end{displaymath}

\textrm{mguto} is a modified version of \textrm{mgu}: basically it does the same thing, but makes sure the user-given type (the second argument) is more specific or equally general than the inferred type (the first argument).

However, the HM version of $\mu \mathrm{E}$ is not as straightforward as the stlc style one. If typechecking $e$ gives us a plain type variable, then we'll be left in the same embarrassing situation of second order type unification. Only when $e$ is inferred to have a type of the form $\mu \alpha. M$ won't we need a type annotation.

\begin{displaymath}
  \frac
    {\Gamma \backslash \Delta \vdash e : \mu \alpha. M, S}
    {\Gamma \backslash \Delta \vdash \unfoldi e : M[\alpha := \mu \alpha. M], S}
  \mu \mathrm{E^I}
\end{displaymath}

\begin{displaymath}
  \frac
    {\Gamma \backslash \Delta \vdash e : \tau, S_0 \qquad S_1 = \mguto{\tau}{\mu \alpha. M} }
    {\Gamma \backslash \Delta \vdash \unfold{\mu \alpha. M}{e} : S_1 (M[\alpha := \mu \alpha. M]), S_1 S_0}
  \mu \mathrm{E^E}
\end{displaymath}

Then we may have the ``constructors'' for \textrm{Nat}:

\begin{displaymath}
\mathrm{Zero} : \mathrm{Nat} = \fold{\mathrm{Nat}}{(\inl \unit)}
\end{displaymath}
\begin{displaymath}
\mathrm{Succ} : \mathrm{Nat} \multimap \mathrm{Nat} = \lambda n. \fold{\mathrm{Nat}}{(\inr n)}
\end{displaymath}

\subsection{The !-modality and General Recursion}

Having a term typed $\bang \tau$ is to have infinite terms typed $\tau$, each with the same value. Thus instead of making it a predefined type, we may as well try to take the advantage of lazy evaluation and define it as an infinite stream: $\bang \tau = \mu \alpha. \tau \otimes \alpha$. We will prove its correspondence to the !-modality in logic shortly, however, in programming, we must use this type with care, since nothing prevents this infinite stream to contain different values.

With the newly defined !-modality, it's now possible to add general recursion to our language by introducting a fixpoint operator, with the following typing rule:
\begin{displaymath}
  \frac
    {\tau' \leftarrow \newvar \qquad
    \cdot, p : \bang \tau' \backslash \Delta \vdash e : \tau, S_0 \qquad
    S_1 = \mgu{\tau}{\tau'}}
    {\Gamma \backslash \Gamma - p \vdash \fix{p}{e} : S_1 \tau, S_1 S_0}
  \mathrm{Fix}
\end{displaymath}

The variable introduced by \textsf{fix} has a $!-$ type, since the function may recursively call itself arbitrarily many times. No other linear hypothesis is allowed to appear in the context of the body part since the computation may recurse more than once, duplicating hypotheses implicitly.

As an example, we can define natural number addition using \textsf{fix}:
\begin{align*}
  \mathrm{Plus} : \operatorname{ } & \mathrm{Nat} \multimap \mathrm{Nat} \multimap \mathrm{Nat}\\
\text{= } & \fix{(ps : \bang (\mathrm{Nat} \multimap \mathrm{Nat} \multimap \mathrm{Nat}))}{} \\
  & \lambda (m : \mathrm{Nat}). \lambda (n : \mathrm{Nat}). \\
  & \textsf{case} ~ m \textsf{ of } \\
  & ~ \textsf{inl} ~ s \Rightarrow n ~ | \\
  & ~ \textsf{inr} ~ m' \Rightarrow \letin{p \otimes ps' = ps}{\textrm{Succ} ~ (p ~ m' ~ n)}
\end{align*}

Now, to convince the reader that the $!-$ type synonym defined above is indeed the !-modality in affine logic, we prove the following two rules, where $\bang \Gamma$ suggests that all hypotheses in $\Gamma$ have a $!-$ type:

\begin{displaymath}
  \frac
    {\Gamma, A \vdash B}
    {\Gamma, ! A \vdash B}
  (1)
  \qquad
  \frac
    {\bang \Gamma \vdash A}
    {\bang \Gamma \vdash ! A}
  (2)
\end{displaymath}

The proof for rule (1) is trivial:
\begin{displaymath}
  \displaystyle \frac
    { \displaystyle \frac 
        {}
        {\Gamma, as : \bang \alpha \backslash \Gamma \vdash as : \bang \alpha} 
      \mathrm{Var} \qquad
      \displaystyle \frac
        {\Gamma, a : \alpha \backslash \cdot \vdash b : \beta}
        {\Gamma, a : \alpha, as' : \bang \alpha \backslash \cdot \vdash b : \beta}
      \mathrm{Weaken}
    }
    {\Gamma, as : \bang \alpha \backslash \cdot \vdash \letin{a \otimes as' = as}{b} : \beta}
  \otimes \mathrm{E}
\end{displaymath}
where $\Gamma \backslash \cdot \vdash e : \tau$ corresponds to $\Gamma \vdash \tau$ in affine logic. Also, while in type inference the usage of the weakening rule is implicit, here when doing reasoning it's written explicitly.

We can only prove the admissibility of rule (2). We start by defining a polymorphic function to contract $!-$ types, which does exactly the same thing as Hilbert when infinitely many passengers entered his hotel. Type annotations are added as much as possible to help understanding:
\begin{align*}
\textrm{Dup!} : \bang \tau \multimap \bang \tau \otimes \bang \tau 
\text{= } & \textsf{fix}(fs : \bang (\bang \tau \multimap \bang \tau \otimes \bang \tau)). \lambda (xs : \bang \tau). \\
  & \letin{a \otimes xs' = \unfold{\bang \tau}{xs}}{} \\
  & \letin{b \otimes xs'' = \unfold{\bang \tau}{xs'}}{} \\
  & \letin{f \otimes fs' = \unfold{\bang (\bang \tau \multimap \bang \tau \otimes \bang \tau)}{fs}}{} \\
  & \letin{as \otimes bs = f ~ xs''}{} \\
  & \fold{\bang \tau}{(a \otimes as)} \otimes \fold{\bang \tau}{(b \otimes bs)}
\end{align*}

Though our language doesn't have a mechanism of termination check, the reader may verify manually that the program indeed terminates when only finite length of the proof term is evaluated, thus we consider this function a valid proof. So is the rest of this proof.

We first consider the case where $!\Gamma$ contains exactly one hypothesis $! A$. We can eliminate the context by lambda abstraction:
\begin{displaymath}
  \frac
    {\cdot, as : \bang \alpha \backslash \cdot \vdash b : \beta}
    {\cdot \backslash \cdot \vdash \lambda as. b : \bang \alpha \multimap \beta}
  \multimap \mathrm{I}
\end{displaymath}

We assign a name $E$ to this term, and define
\begin{align*}
F : \bang \alpha \multimap \bang \beta 
\text{= } & \textsf{fix }(fs : \bang (\bang \alpha \multimap \bang \beta). \lambda (as : \bang \alpha). \\
  & \letin{as' \otimes as'' = \textrm{Dup!} ~ as}{} \\
  & \letin{f \otimes fs' = \unfold{\bang (\bang \alpha \multimap \bang \beta)}{fs}}{} \\
  & \fold{\bang \beta}{(E ~ as' \otimes f ~ as'')}
\end{align*}

Then by application, we can again put $\bang \alpha$ back into the context:
\begin{displaymath}
  \displaystyle \frac
    { \displaystyle \frac
        {F : \bang \alpha \multimap \bang \beta}
        {\cdot, as : \bang \alpha \backslash \cdot, as : \bang \alpha \vdash F : \bang \alpha \multimap \bang \beta}
      \mathrm{Intro}
      \qquad
      \displaystyle \frac
        {}
        {\cdot, as : \bang \alpha \backslash \cdot \vdash as : \bang \alpha}
      \mathrm{Var}
    }
    {\cdot, as : \bang \alpha \backslash \cdot \vdash F ~ as : \bang \beta}
  \multimap \mathrm{E}
\end{displaymath}

The case where $!\Gamma$ consists of multiple linear hypotheses can also be proved using the the same approach.

We left the case where $!\Gamma = \cdot$ to the last, because it's a rather common pattern and is thus very useful.

\begin{displaymath}
  \displaystyle \frac 
    { \displaystyle \frac 
        { \displaystyle \frac 
            { \displaystyle \frac 
                {E : \tau} 
                {xs : \bang \tau \backslash xs : \bang \tau \vdash E : \tau}
              \mathrm{Intro}
              \qquad
              \displaystyle \frac 
                {} 
                {xs : \bang \tau \backslash \cdot \vdash xs : \bang \tau}
              \mathrm{Var}
            }
            {xs : \bang \tau \backslash \cdot \vdash E \otimes xs : \tau \otimes \bang \tau} 
          \otimes \mathrm{I}
        }
        {xs : \bang \tau \backslash \cdot \vdash \fold{\bang \tau}{(E \otimes xs)} : \bang \tau} 
      \mu \mathrm{I}
    }
    {\cdot \backslash \cdot \vdash \fix{xs}{\fold{\bang \tau}{(E \otimes xs)}} : \bang \tau} 
  \mathrm{Fix}
\end{displaymath}

Combine it with the $\mathrm{Intro}$ rule and we're able to introduce infinite $E$s in any context if we have a globally defined term $E$. Hence we may create a syntatic sugar as the introduction rule for the !-modality:
\begin{displaymath}
  \frac
    {E : \tau \qquad  \tau' \leftarrow \inst{\tau}}
    {\Gamma \backslash \Gamma \vdash \bang E : \bang \tau', \emptyset}
    !\mathrm{I}
\end{displaymath}
which desugars to the derivation above. This rule again reveals the nature of global terms that they reflect constant truth logically and (often) compile to static code when implemented.

\section{Conclusion}

In this article, we showed how to combine the typing rules of an affine type system with Hindley-Milner style type inference, by generalizing algorithm W, which is substitution-based and highly inefficient. In practice, algorithm J enjoys a nearly linear time complexity and thus is more preferrable, whose rules may be easily deduced from that of algorithm W. We also discussed the possibility to represent the !-modality as an infinite stream in a lazy language, and proved its correctness.

\section*{Appendix}

Here is a Haskell implementation of the most general unifier algorithm.

\begin{lstlisting}
import qualified Data.Map as Map

data TConst = TOne | TZero deriving (Show, Eq)
data TOp = TTensor | TPlus | TWith deriving (Show, Eq)

data Type 
  = TConst TConst
  | TOp TOp Type Type
  | TMu String Type
  | TVar String
  deriving (Show, Eq)

type Substitution = Map.Map String Type

compose :: Substitution -> Substitution -> Substitution
compose s1 s2 = fmap (subst s1) s2 `Map.union` s1

free :: Type -> [String]
free (TConst _)    = []
free (TOp _ t1 t2) = free t1 ++ free t2
free (TMu x t)     = filter (/= x) (free t)
free (TVar x)      = [x]

subst :: Substitution -> Type -> Type
subst _ (TConst c)     = TConst c
subst s (TOp op t1 t2) = TOp op (subst s t1) (subst s t2)
subst s (TMu x t)      = TMu x (subst (Map.delete x s) t)
subst s (TVar x)       = case Map.lookup x s of
  Just t  -> t
  Nothing -> TVar x

rename :: String -> String -> Type -> Type
rename v v' = subst (Map.singleton v (TVar v'))

mguPrim :: Bool -> Type -> Type -> Either String Substitution
mguPrim _ (TConst c1) (TConst c2)
  | c1 == c2  = pure Map.empty
  | otherwise = Left $ "can't unify " ++ show c1 ++ " and " ++ show c2
mguPrim b (TOp op t1 t2) (TOp op' t1' t2')
  | op == op' = do
      s1 <- mguPrim b t1 t1'
      s2 <- mguPrim b (subst s1 t2) (subst s1 t2')
      pure (s2 `compose` s1)
  | otherwise = Left $ "can't unify " ++ show (TOp op t1 t2) ++ " and " ++ show (TOp op' t1' t2')
mguPrim b (TMu v t) (TMu v' t')
  | v == v'   = mguPrim b t t'
  | otherwise = mguPrim b t (rename v' v t')
mguPrim _ (TVar v) t
  | t == TVar v     = pure Map.empty
  | v `elem` free t = Left $ "occurs check failed: " ++ v ++ " in " ++ show t
  | otherwise       = pure (Map.singleton v t)
mguPrim b t (TVar v) = if b
  then mguPrim b (TVar v) t
  else Left $ "can't unify " ++ show t ++ " to given type variable " ++ v ++ " which is more general"
mguPrim _ a b = Left $ "can't unify " ++ show a ++ " and " ++ show b

mgu :: Type -> Type -> Either String Substitution
mgu = mguPrim True

mguTo :: Type -> Type -> Either String Substitution
mguTo = mguPrim False
\end{lstlisting}

\end{document}